%
%
%
%

%
%
%
%

\magnification=1200
\raggedbottom
\voffset6truemm
\nopagenumbers
\def\II{{\rm 1\!\hskip-1pt I}}
\def\cstok#1{\leavevmode\thinspace\hbox{\vrule\vtop{\vbox{\hrule\kern1pt
\hbox{\vphantom{\tt/}\thinspace{\tt#1}\thinspace}}
\kern1pt\hrule}\vrule}\thinspace}
\headline={\ifnum\pageno=1\hfill\else
\sf I. G. Avramidi and G. Esposito: Lack of ellipticity in 
Euclidean quantum gravity
\hfill\rm\folio \fi}

\font\unifnt=cmss10 scaled \magstep1
\font\titlefnt=cmbx12 scaled \magstep2
\font\sectionfnt=cmbx10 scaled \magstep1
\font\authorfnt=cmr12 scaled \magstep1

\font\sf=cmss10

{
\null
\vskip-3truecm
\hskip7truecm
{\hrulefill}
\par
\vskip-4truemm
\par
\hskip7truecm
{\hrulefill}
\par
\vskip5mm
\par
\hskip7truecm
{{\unifnt University of Greifswald (August, 1997)}}
\vskip4mm
\par
\hskip7truecm
{\hrulefill}
\par
\vskip-4truemm
\par
\hskip7truecm
{\hrulefill}
\par
\par
\bigskip
\hskip7truecm{\unifnt DSF 97/43}
\bigskip 
\hskip7truecm{\ hep-th/9708163}
\bigskip
\smallskip
\hskip7truecm{\ Submitted to:}{\ \sf Class. Quant. Grav.}
\vfill

\centerline {\titlefnt Lack of strong ellipticity}
\medskip
\centerline{\titlefnt in Euclidean quantum 
gravity}
\vskip 1cm
\centerline {{\authorfnt Ivan G. Avramidi},$^{1}$
\footnote{$^*$}{{\rm
On leave of absence from Research Institute
for Physics, Rostov State University, Stachki 194,
344104 Rostov-on-Don, Russia. E-mail: 
avramidi@rz.uni-greifswald.de}} and 
{\authorfnt Giampiero Esposito}$^{2,3}$}
\vskip 1cm
\noindent
\it
\centerline{
${ }^{1}$Department of Mathematics, University of Greifswald,}
\centerline{
Jahnstr. 15a, D-17487 Greifswald, Germany}
\vskip 0.3cm
\noindent
\centerline{
${ }^{2}$Istituto Nazionale di Fisica Nucleare, Sezione di
Napoli,}
\centerline{
Mostra d'Oltremare Padiglione 20, 80125 Napoli, Italy}
\vskip 0.3cm
\noindent
\centerline{
${ }^{3}$Dipartimento di Scienze Fisiche,}
\centerline{ Mostra d'Oltremare
Padiglione 19, 80125 Napoli, Italy}
\vskip 0.3cm
\noindent
\vskip 1cm
\noindent
\vfill
\noindent
\rm
{\par
\narrower\noindent
{\bf Abstract}.
Recent work in Euclidean quantum gravity has studied 
boundary conditions which are completely invariant under
infinitesimal diffeomorphisms on metric perturbations. On
using the de Donder gauge-averaging functional, this scheme
leads to both normal and tangential derivatives in the
boundary conditions. In the present paper, it is proved that
the corresponding boundary value problem fails to be strongly
elliptic. The result raises deep interpretative issues for
Euclidean quantum gravity on manifolds with boundary.
\par
}
\eject

\leftline{\sectionfnt 1. Introduction}
\vskip 0.3cm
\noindent
The problem of boundary conditions has always been of crucial
importance for a thorough understanding of the quantized
gravitational field and for the attempts to develop a quantum
theory of the universe [1, 2]. In particular, in recent years, 
many efforts have been produced to study mixed boundary 
conditions in the one-loop semiclassical approximation for
pure gravity. 

As is well known, the one-loop contribution to the effective action 
of a gauge theory (with closed algebra of gauge generators) is
determined by the functional determinats of 
some differential operators [3]
$$
\Gamma_{(1)}={1\over 2}\log{\rm Det}\, F -\log {\rm Det}\, D
\eqno (1.1)
$$
where $F$ is the gauge field operator determined by the second variation
of the classical action with respect to the background 
fields and suitable gauge-averaging terms,
and $D$ is the ghost operator determined by the generators
of gauge transformations and the gauge-averaging functional.
Using the condensed notation of DeWitt [3], one can write them in the form
$$
F_{ik}=S_{,ik}+E_{im}R^m{}_\alpha\gamma^{\alpha\beta}R^n{}_\beta E_{nk}
\eqno (1.2)
$$
$$
D_{\alpha\beta}=R^i{}_\alpha E_{ik}R^k{}_\beta
\eqno (1.3)
$$
where $S$ is the action functional, $E_{ik}$ 
is the metric in the configuration space,
$R^i{}_\alpha$ are the generators of gauge 
transformations, $\gamma^{\alpha\beta}$
is a constant ultralocal matrix, comma denotes 
the functional derivative and a combined summation-integration
over the discrete and continuous indices is assumed.

This expression is believed to be a covariant (gauge invariant) functional.
By using the so called minimal Landau-DeWitt gauge
(which is also called de Donder-Fock gauge in the case of gravity)
it is possible to make the gauge field 
operator $F$ of Laplace type (or minimal, in the physical
terminology [3]; see section 2).

The functional determinants are well defined 
only for {\it elliptic} differential 
operators. Therefore, in the case of incomplete manifolds $M$, i.e.  
with a boundary $\partial M$, the differential operators
should be supplied with some suitable boundary conditions, which
make them self-adjoint and elliptic, say,
$$
B_F h\big|_{\partial M}=0 \qquad B_D\,\varphi\big|_{\partial M}=0
\eqno (1.4)
$$
where $h\in C^\infty(T^*M\otimes T^*M,M)$ and $\varphi\in C^\infty(TM,M)$ 
are the metric perturbations and ghost fields, respectively.
On the other hand, such boundary conditions should be gauge invariant, i.e.
invariant under the gravitational (infinitesimal) gauge transformations
$$
\delta_\xi h={\cal L}_\xi h \qquad \delta_\xi\varphi={\cal L}_\xi\varphi
\eqno (1.5)
$$
where $\xi\in C^\infty(TM,M)$ is an arbitrary 
vector field and ${\cal L}_\xi$ is the Lie derivative along $\xi$.

In the scheme proposed first by Barvinsky [4] the gauge invariant boundary
conditions for quantum gravity have the form, in the de Donder gauge,
$$
h_{ij}\big|_{\partial M}=0 \qquad
E^{abcd}\nabla_{b}h_{cd}\big|_{\partial M}=0
\eqno (1.6)
$$
$$
\varphi_{a}\big|_{\partial M}=0
\eqno (1.7)
$$
where $E^{abcd}$ is the local metric in the space of 
metric perturbations. At a deeper level, such boundary conditions
are BRST invariant [5].
On separating the normal derivative, the boundary 
conditions (1.6) turn out to be an extension of
the generalized boundary value problem, which includes
(unlike the usual Dirichlet or Neumann conditions)
both the normal derivative and the tangential derivatives 
(see section 2).

In [6] it was proved that the operator $F$ 
for gravity, with the boundary conditions
(1.6), is symmetric. 
Moreover, heat-kernel asymptotics with tangential
derivatives in the boundary conditions is now receiving 
consideration for the first time [2, 7--10], although the 
physical motivation was already clear from the work in 
[4] and [11]. Our paper, however, investigates a foundational
issue whose consideration comes before any attempt to perform
lengthy calculations. For this purpose, section 2 defines and
studies strong ellipticity for generalized boundary value
problems involving operators of Laplace type. The crucial 
step, i.e. the Euclidean quantum gravity analysis, is
undertaken in detail in section 3. Concluding remarks are
presented in section 4, and relevant background material is
described in the appendix.

\bigskip
\bigskip
\leftline{\sectionfnt 2. Strong ellipticity of the generalized boundary value problem}
\vskip 0.3cm
\noindent
As a first step in our investigation, we are now going to
study when a Laplace type operator, subject
to generalized boundary conditions (see below), satisfies the
Lopatinski-Shapiro strong ellipticity condition [12, 13].

Let $V$ be a vector bundle over a compact Riemannian manifold $M$
with positive-definite metric $g$ and a smooth boundary $\partial M$, 
and let $C^\infty(V,M)$ be the space of smooth
sections of the bundle $V$. Using a Hermitian metric $E$ and the Riemannian
volume element on $M$, the dual bundle $V^*$ is naturally 
identified with $V$ and 
a natural $L^2$ inner product is defined. 
The Hilbert space $L^2(V,M)$ is then defined
to be the completion of the space $C^\infty(V,M)$. 

An operator of Laplace type, say $F$, is a map [12]
$$
F: C^{\infty}(V,M) \longrightarrow C^{\infty}(V,M)
\eqno (2.1)
$$
which can be expressed in the form
$$
F=-g^{ab}\nabla_{a}^{V}\nabla_{b}^{V}+Q
\eqno (2.2)
$$
where $\nabla^{V}$ is the connection on $V$ and $Q$ is
a self-adjoint endomorphism of $V$. 
The adjoint operator $\bar F$ is defined using the $L^2$ inner product, 
i.e. $(\bar F \varphi,\psi)=(\varphi,F\psi)$.

The task, in general,  
is to prove that the Laplace type operator with
suitable boundary conditions is an {\it essentially self-adjoint}
and {\it elliptic} operator, which means that it is: i) {\it symmetric}, i.e.
$(F\varphi,\psi)=(\varphi,F\psi), {\rm for\  all}\ \varphi, \psi$; ii) 
{\it strongly elliptic} and iii) there exists
a unique {\it self-adjoint extension} of $F$.
We are going to study the first and the second question but not the last one.

The generalized boundary conditions that guarantee 
the symmetry of the operator $F$ are [6, 7]
$$
\Pi\, \varphi \big|_{\partial M}=0
\eqno (2.3)
$$
$$
(\II-{\overline \Pi})(\nabla_{0}+\Lambda)
\varphi \big|_{\partial M}=0
\eqno (2.4)
$$
where $\Pi$ is a projector, ${\overline \Pi}$ is the 
dual projector ${\overline \Pi} \equiv E^{-1}\Pi^{\dag}E$
and $\Lambda$ is a self-adjoint tangential operator
of first order. It can be always put in the form
$$
\Lambda=(\II-{\overline \Pi})\left[
{1\over 2}\Bigr(\gamma^{i}{\widehat \nabla}_{i}
+{\widehat \nabla}_{i}\gamma^{i}\Bigr)+S
\right](\II-\Pi)
\eqno (2.5)
$$
where the matrices $\gamma^i$ and $S$ satisfy the conditions
$$
{\overline \gamma}^{i}\equiv E^{-1}\gamma^{i\dag}E=-\gamma^{i}
\qquad
{\overline S} \equiv E^{-1}S^{\dag}E=S
\eqno (2.6)
$$
$$
{\overline \Pi}\gamma^i=\gamma^i\Pi=0
\eqno (2.7)
$$
$$
{\overline \Pi}S=S\Pi=0.
\eqno (2.8)
$$

To begin, note that the leading symbol of
the operator $F$ reads [12]
$$
\sigma_{L}(F;x,\xi)={\mid \xi \mid}^{2} \equiv g^{\mu \nu}
(x)\xi_{\mu} \xi_{\nu} \; \II 
\eqno (2.9)
$$
where $\xi \in T^{*}(M)$ is a cotangent vector and $\II$ is the
identity endomorphism of $V$. Of course, for a positive-definite
non-singular metric the leading symbol is non-degenerate for
$\xi \not = 0$. Moreover, for a complex $\lambda$ which does
not lie on the positive real axis, $\lambda \in {\bf C}
-{\bf R}_{+}$, one has
$$
{\rm det} (\sigma_{L}(F;x,\xi)-\lambda)
=({\mid \xi \mid}^{2}-\lambda)^{{\rm dim}V} \not = 0.
\eqno (2.10)
$$
This equals zero only for $\xi=\lambda=0$. Thus, the leading symbol
of the operator $F$ is elliptic.

To formulate the strong ellipticity condition for the boundary 
value problem (see, e.g. [12] and [13]) 
we introduce first some notation (we stress that 
we consider only second-order operators). Let
$$
W = W_{0} \oplus W_{1}
\eqno (2.11)
$$
with 
$$
W_{0}=\left \{ \varphi \mid_{\partial M} \right \}
\; \; \; \; 
W_{1}=\left \{ \nabla_{0} \varphi \mid_{\partial M}
\right \}
\eqno (2.12)
$$
be the bundle of boundary data. Let the operator
$$
K: C^{\infty}(V,M) \rightarrow C^{\infty}(W,\partial M)
\eqno (2.13)
$$
be the boundary data map
$$
K \varphi=\pmatrix{\varphi \mid_{\partial M}\cr 
\nabla_{0}\varphi \mid_{\partial M}\cr}.
\eqno (2.14)
$$
Moreover, we consider an auxiliary vector bundle over 
$\partial M$, $W'=W_{0}' \oplus W_{1}'$, having the same
dimension as $V$, and a tangential differential operator on
$\partial M$, say $B: C^{\infty}(W,\partial M) \rightarrow
C^{\infty}(W',\partial M)$, written as a matrix
$$
B=\pmatrix{B_{00}& B_{01}\cr B_{10}& B_{11}\cr}.
\eqno (2.15)
$$
Further assume that (hereafter, ``ord" denotes the order
of the differential operator)
$$
{\rm ord}(B_{ij}) \leq i-j
\eqno (2.16)
$$
and define the graded order of $W_{j}'$ to be j:
$$
{\rm ord}(\varphi \mid_{\partial M})=0 \; \; \; \;
{\rm ord}(\nabla_{0}\varphi \mid_{\partial M})=1
\eqno (2.17)
$$
and finally the graded leading symbol of $B$ by
$$
\sigma_{g}(B_{ij})=\cases{
\sigma_{L}(B_{ij}) \; & {\rm if} \  ${\rm ord}(B_{ij})=i-j$ \cr
0 \; &{\rm if} \ ${\rm ord}(B_{ij}) < i-j$ \cr}.
\eqno (2.18)
$$
The boundary conditions can then be written in the form
$$
{\cal B}\varphi=0
\eqno (2.19)
$$
where $\cal B$ is the boundary operator defined by
$$
{\cal B}\varphi \equiv B K.
\eqno (2.20)
$$

For the generalized boundary conditions (2.3) and (2.4) we set
again $W'=W_{0}' \oplus W_{1}'$, where
$$
W_{0}' \equiv \{\Pi \varphi|_{\partial M}\} \qquad 
W_{1}' \equiv \{(\II-{\overline \Pi})\nabla_0\varphi|_{\partial M}\}.
\eqno (2.21)
$$
The operator $B$ is easily found to be (see (2.3) and (2.4))
$$
B=\pmatrix{\Pi & 0 \cr 
\Lambda & (\II-{\overline \Pi})\cr}
\eqno (2.22)
$$
with graded leading symbol
$$
\sigma_{g}(B)=\pmatrix{\Pi & 0 \cr iT &
(\II-{\overline \Pi})\cr}
\eqno (2.23)
$$
where
$$
T \equiv (\II-{\overline \Pi})\gamma^{j}\zeta_{j}
(\II-\Pi) = \gamma^{j}\zeta_{j}
\eqno (2.24)
$$
$\zeta_{i} \in T^{*}(\partial M)$
being a cotangent vector on the boundary.

To define the strong ellipticity condition [12], we take the
leading symbol 
\break
$\sigma_{L}(F;{\hat x},r,\zeta,\omega)$ of the
operator $F$, replace $\omega$ by $-i \partial_{r}$ and consider
the following ordinary differential equation:
$$
[\sigma_{L}(F;{\hat x},r,\zeta,-i \partial_{r})-\lambda]
\varphi=0.
\eqno (2.25)
$$
A second-order operator $F$ with the boundary conditions defined
by the operator $B$ is said to be {\it strongly elliptic} if
there exists a unique solution of equation (2.25) for
$(\zeta,\lambda) \not = (0,0)$ subject to the asymptotic
condition
$$
\lim_{r \to \infty}\varphi=0
\eqno (2.26)
$$
and to the boundary condition
$$
\sigma_{g}(B)K \varphi=\psi'
\eqno (2.27)
$$
for any $\psi' \in W'$.

For an operator of Laplace type, the equation (2.25) 
takes the form
$$
\Bigr(-\partial_{r}^{2}+{{\zeta}}^{2}
-\lambda \Bigr)\varphi=0
\eqno (2.28)
$$
where ${{\zeta}}^{2} \equiv \gamma^{ij}(\hat x)
\zeta_{i}\zeta_{j}$. The general solution of (2.28) satisfying the
asymptotic condition (2.26) reads
$$
\varphi=\chi \exp(-\mu r)
\eqno (2.29)
$$
where $\mu=\sqrt{{{\zeta}}^{2}-\lambda}$. Since 
$(\zeta,\lambda) \not = (0,0)$, and bearing in mind that
$\lambda \in {\bf C}-{\bf R}_{+}$, one can always choose
${\rm Re}(\mu)>0$. Thus, the question of 
strong ellipticity for Laplace
type operators is reduced to the invertibility of the equations
$$
\pmatrix{\Pi & 0 \cr iT & (\II-{\overline \Pi})\cr}
\pmatrix{\chi \cr -\mu \chi \cr}
=\pmatrix{\Pi \psi_{0} \cr -\mu(\II -{\overline \Pi})
\psi_{0}\cr}
\eqno (2.30)
$$
which can be rewritten in the form
$$ \eqalignno{
\; & \pmatrix{\II & 0 \cr 
\mu(\II-{\overline \Pi})&
(\II-{\overline \Pi})\mu -iT\cr}
\pmatrix{\Pi \chi \cr 
(\II- \Pi)\chi \cr} \cr 
&=\pmatrix{\Pi \psi_{0} \cr
\mu (\II-{\overline \Pi})\psi_{0} \cr}
&(2.31)\cr}
$$
and can be transformed into
$$
\pmatrix{\II & 0 \cr 
0 & \beta\mu -iT\cr}
\pmatrix{\Pi \chi \cr (\II-\Pi)\chi \cr}
=\pmatrix{\Pi \psi_{0} \cr 
\mu\beta(\II-\Pi) \psi_{0}\cr}
\eqno (2.32)
$$
where 
$$
\beta \equiv (\II-\bar\Pi)(\II-\Pi)+\bar\Pi\varepsilon\Pi
\eqno (2.33)
$$
and $\varepsilon$ is an {\it arbitrary} self-adjoint 
matrix, $\bar\varepsilon=\varepsilon$.

If this equation has a unique solution for any $\psi_{0} \in W_{0}$, 
then the boundary value problem is strongly elliptic.
In other words, the boundary value problem is strongly elliptic 
if the matrix on the left hand side of equation (2.32)
is invertible, which is equivalent to the non-degeneracy 
of the matrix $[\beta\mu-iT]$, i.e.
$$
\det\pmatrix{\II & 0 \cr 
0 & \beta\mu -iT\cr}=\det[\beta\mu-iT]\ne 0
\eqno (2.34)
$$
for any $(\zeta,\lambda)\ne (0,0)$ and $\lambda\in {\bf C}-{\bf R}_+$.
If this condition is satisfied, the solution of 
equation (2.32) is given by
$$
\pmatrix{\Pi \chi \cr (\II-\Pi)\chi \cr}
=\pmatrix{\II & 0 \cr 
0 & (\beta\mu -iT)^{-1}\cr}
\pmatrix{\Pi \psi_{0} \cr 
\mu\beta(\II-\Pi) \psi_{0}\cr}.
\eqno (2.35)
$$
Note that the matrix $\beta$ is self-adjoint, $\bar\beta=\beta$.
It is very convenient to choose $\varepsilon$
in such a way that the matrix $\beta$ becomes non-degenerate, 
$\det\,\beta\ne 0$. One can then define
$$
Y^{i} \equiv \beta^{-1}\gamma^{i}
\eqno (2.36)
$$
and
$$
X \equiv \beta^{-1}T = Y^{i}\zeta_{i} .
\eqno (2.37)
$$
Since the $\gamma^{i}$ are anti-self-adjoint,
the matrices $Y^{i}$ and $X$ are also anti-self-adjoint
$$
{\overline Y}^{i}=-\beta Y^{i}\beta^{-1}
\eqno (2.38)
$$
$$
{\overline X}=-\beta X \beta^{-1}.
\eqno (2.39)
$$

If the matrix $\beta$ is non-degenerate, 
the solution of equation (2.32) takes the form
$$
\pmatrix{\Pi \chi \cr (\II-\Pi)\chi \cr}
=\pmatrix{\II & 0 \cr 
0 & (\mu -iX)^{-1}\cr}
\pmatrix{\Pi \psi_{0} \cr 
\mu(\II-\Pi) \psi_{0}\cr}
\eqno (2.40)
$$
and the condition of strong ellipticity  
reduces to the non-degeneracy of the
matrix $(\II \mu -i X)$, i.e.
$$
\det[\II\mu-iX]={\rm det} \Bigr[\II \sqrt{{{\zeta}}^{2}-\lambda}
-i \beta^{-1}\gamma^{j}\zeta_{j}\Bigr]
\not = 0
\eqno (2.41)
$$
for $(\zeta,\lambda) \not = (0,0)$ and $\lambda \in 
{\bf C}-{\bf R}_{+}$. 

Moreover, noting that
$$
(\II \mu -iX)(\II \mu +iX)=\II \mu^{2}+X^{2}
\eqno (2.42)
$$
we obtain eventually the strong ellipticity condition
in the most convenient form:
$$
{\rm det} \Bigr[\II(-\lambda+\zeta^{2})+X^{2} \Bigr]
\not = 0
\eqno (2.43)
$$
or
$$
{\rm det} \Bigr[-\II \lambda+(\II \gamma^{jk}
+Y^{(j}Y^{k)})\zeta_{j}\zeta_{k}\Bigr] \not = 0
\eqno (2.44)
$$
for $(\zeta,\lambda) \not = (0,0), \lambda \not \in {\bf R}_{+}$.

This means that, for the boundary value problem to be strongly
elliptic, the eigenvalues of the matrix $X^{2}$ should be {\it real}
and larger than $-\zeta^{2}$, i.e.
$$
{\rm Im}(X^{2})=0 \; \; \; \; 
{\rm Re}(X^{2}+ \II \zeta^{2}) > 0
\eqno (2.45)
$$
for any cotangent vector $\zeta_{j}$.

\bigskip
\bigskip
\leftline{\sectionfnt 3. Lack of strong ellipticity in Euclidean quantum gravity}
\vskip 0.3cm
\noindent
Now we study in detail the generalized boundary conditions (1.6)
in Euclidean quantum gravity. The vector  bundle $V$ is here 
the vector bundle of symmetric rank-two
tensors on $M$: $V=T^*M\otimes T^*M$. 
This bundle has a connection [9]
$$
\omega_{e,ab}^{\; \; \; \; \; \; cd}
=-2\Gamma_{\; \; \; e(a}^{(c} 
\delta_{\; \; \; b)}^{d)}
\eqno (3.1)
$$
and a curvature
$$
\Omega_{ef,ab}^{\; \; \; \; \; \; \; \; cd}
=-2R_{ef(a}^{\; \; \; \; \; \; \; (c}  
\delta_{b)}^{\; \; \; d)}
\eqno (3.2)
$$
and its metric is defined by the equation
$$
E^{ab \; cd} \equiv g^{a(c} g^{d)b}
-{1\over 2}g^{ab}g^{cd}.
\eqno (3.3)
$$
Note that 
$$
E_{ab \; cd}^{-1} \equiv g_{a(c} g_{d)b}
-{1\over (m-2)}g_{ab}g_{cd}
\eqno (3.4)
$$
and hence this metric is not well defined for $m=2$. The
corresponding graviton operator $F$ in the covariant de Donder type minimal 
gauge is then of Laplace type (2.2) , with a ``potential
term" constructed from the Riemann curvature tensor [9].

On separating the normal derivative in the boundary conditions
(1.6) and introducing the tensor 
$$
q_{ab} \equiv g_{ab}-N_{a}N_{b}
\eqno (3.5)
$$
we find the boundary operator $\cal B$ exactly as
described in the previous section, with the following matrices [9]:
$$
\II=\II_{ab}^{\; \; \; cd}
\equiv \delta_{(a}^{c} \delta_{b)}^{d}
\eqno (3.6)
$$
$$
\Pi=\Pi_{ab}^{\; \; \; cd} \equiv q_{(a}^{c} \; q_{b)}^{d}
\eqno (3.7)
$$
$$
\Gamma^i\equiv (E\gamma^{i})_{ab}^{\; \; \; cd} \equiv
-N_{a}N_{b}e^{i(c} N^{d)}
+N_{(a} e_{b)}^{i} N^{c} N^{d}
\eqno (3.8)
$$
$$
(ES)_{ab}^{\; \; \; cd} \equiv -N_{a}N_{b}N^{c}N^{d}K
+2N_{(a} e_{b)}^{i} e^{j(c} N^{d)}[K_{ij}
+\gamma_{ij}K].
\eqno (3.9)
$$
The matrices $\gamma^{i}$ are easily computed from the above
equations
$$ \eqalignno{ 
\gamma_{\; ab}^{i \; \; \; cd} & \equiv 
-{(m-3)\over (m-2)} N_{a}N_{b}e^{i(c} N^{d)}
+N_{(a} e_{b)}^{i}N^{c}N^{d} \cr
& +{1\over (m-2)}q_{ab}e^{i(c} N^{d)}.
&(3.10)\cr}
$$
It is easily seen that the matrices $\gamma^{i}$ are anti-self-adjoint
and the matrix $S$ is self-adjoint, and that the conditions (2.6)--(2.8)
are satisfied.

We now introduce further projectors
$$
\kappa \equiv {1\over (m-1)}q_{ab}q^{cd}
\eqno (3.11)
$$
$$
\psi \equiv 2N_{(a}q_{b)}^{\; (c}N^{d)}
\eqno (3.12)
$$
$$
\pi \equiv N_{a}N_{b}N^{c}N^{d} .
\eqno (3.13)
$$
The only non-vanishing products among them are
$$
\kappa^{2}=\kappa \; \; \; \; \psi^{2}=\psi
\; \; \; \; \pi^{2}=\pi
\eqno (3.14)
$$
$$
\Pi \kappa=\kappa \Pi=\kappa .
\eqno (3.15)
$$
Moreover [6]
$$
\II=\Pi+\psi+\pi .
\eqno (3.16)
$$
At this stage we consider the following nilpotent matrices:
$$
p_{1} \equiv q_{ab}N^{c}N^{d}
\eqno (3.17)
$$
$$
p_{2} \equiv N_{a}N_{b}q^{cd}
\eqno (3.18)
$$
$$
p_{1}^{2}=p_{2}^{2}=0 .
\eqno (3.19)
$$
The set of matrices $\Pi,\kappa,\psi,\pi,p_{1},p_{2}$ form a
closed algebra. The non-vanishing elements of their 
multiplication table are
$$
\Pi \Pi=\Pi \; \; \; \; \Pi \kappa=\kappa
\; \; \; \; \Pi p_{1}=p_{1}
\eqno (3.20)
$$
$$
\kappa \Pi=\kappa \; \; \; \; \kappa \kappa=\kappa 
\; \; \; \; \kappa p_{1}=p_{1}
\eqno (3.21)
$$
$$
\psi \psi=\psi
\eqno (3.22)
$$
$$
\pi \pi=\pi \; \; \; \; \pi p_{2}=p_{2}
\eqno (3.23)
$$
$$
p_{1}\pi=p_{1} \; \; \; \; 
p_{1}p_{2}=(m-1)\kappa
\eqno (3.24)
$$
$$
p_{2}\Pi=p_{2} \; \; \; \; p_{2}\kappa=p_{2}
\; \; \; \; p_{2}p_{1}=(m-1)\pi .
\eqno (3.25)
$$
Using the metric $E$ we compute the projector 
$$ \eqalignno{
{\overline \Pi} & \equiv E^{-1} \Pi^{T} E 
=\Pi-{(m-1)\over 2(m-2)}\kappa+{1\over 2(m-2)}p_{1}\cr
&+{(m-3)\over 2(m-2)}p_{2}+{(m-1)\over 2(m-2)}\pi .
&(3.26)\cr}
$$
By varying the matrix $\varepsilon$ we can change 
essentially the matrix $\beta$.
The simplest choice is when the matrix $\varepsilon$ 
is proportional to the identity matrix:
$\varepsilon=\II \sigma$. 
Then the matrix $\beta$ defined in (2.33) reads
$$ \eqalignno{
\beta & =\II-(1-\sigma)\Pi
-{\sigma (m-1)\over 2(m-2)}\kappa-{(m-1)\over 2(m-2)}\pi \cr
& -{1\over 2(m-2)}p_{1}+{\sigma(m-3)\over 2(m-2)}p_{2}.
&(3.27)\cr}
$$
By changing the parameter $\sigma$ one can always 
manage to get a non-degenerate matrix $\beta$.
Surprisingly, the matrices $Y^{i}$ defined in (2.36)
do not depend on $\sigma$ and
read, in the gravitational problem,
$$
Y^{i}=-2 N_{a}N_{b}e^{i(c}N^{d)}
+N_{(a}e_{b)}^{i}N^{c}N^{d}.
\eqno (3.28)
$$
Thus, the matrix $X$ defined in (2.37) is
$$
X=-2 p_{3}+p_{4}
\eqno (3.29)
$$
where
$$
p_{3} \equiv N_{a}N_{b}\zeta^{(c}N^{d)}
\eqno (3.30)
$$
$$
p_{4} \equiv N_{(a} \zeta_{b)}N^{c}N^{d}
\eqno (3.31)
$$
and $\zeta_{a} \equiv e_{a}^{i} \zeta_{i}$, so that 
$\zeta_{a}N^{a}=0$. It is important to note 
$$
\Pi X=X\Pi=0.
\eqno (3.32)
$$
Let us now define another projector,
$$
\rho \equiv {2\over \zeta^{2}}N_{(a}\zeta_{b)}
N^{(c} \zeta^{d)} \; \; \; \; 
\rho^{2}=\rho .
\eqno (3.33)
$$
The matrices $p_{3}$ and $p_{4}$ are nilpotent: 
$p_{3}^{2}=p_{4}^{2}=0$, and their products are proportional
to the projectors
$$
p_{3}p_{4}={1\over 2}\zeta^{2}\pi
\eqno (3.34)
$$
$$
p_{4}p_{3}={1\over 2}\zeta^{2}\rho .
\eqno (3.35)
$$
Thus, one finds
$$
X^{2}=-\zeta^{2}(\pi+\rho) .
\eqno (3.36)
$$
Taking into account the orthogonality of the projectors 
$\pi$ and $\rho$: $\pi \rho=\rho \pi=0$, we compute further
$$
X^{2n}=(i {\zeta})^{2n}(\pi+\rho)
\eqno (3.37)
$$
$$
X^{2n+1}=(i {\zeta})^{2n}X .
\eqno (3.38)
$$
Last, since $\pi$ and $\rho$ have unit trace, whilst $p_{3}$
and $p_{4}$ have vanishing trace, we obtain
$$
{\rm tr}(X^{2n})=2 (i {\zeta})^{2n} \; \; \; \; 
{\rm tr}(X^{2n+1})=0 .
\eqno (3.39)
$$
The above properties imply the following theorem:
\vskip 0.3cm
\noindent
{\bf Theorem 3.1} For any function $f$ analytic at the origin one has
$$ \eqalignno{
f(X)&=f(0)[\II-\pi-\rho]+{1\over 2}[f(i\zeta)+f(-i\zeta)]
(\pi+\rho) \cr
&+{1\over 2i \zeta}[f(i \zeta)-f(-i \zeta)]X
&(3.40)\cr}
$$
$$
{\rm tr}f(X)=\left[{m(m+1)\over 2}-2 \right]f(0)
+f(i\zeta)+f(-i\zeta) .
\eqno (3.41)
$$
As a corollary, the eigenvalues of the matrix $X$ are
$$
{\rm spec}\,(X)=\cases{
0 \;& {\rm with} \ {\rm degeneracy} \
$\left[{m(m+1)\over 2}-2 \right]$\cr
i \zeta \ & {\rm with} \  {\rm degeneracy} \  $1$\cr
-i \zeta \ & {\rm with} \ {\rm degeneracy} \ $1$\cr}.
\eqno (3.42)
$$
Thus, the eigenvalues of the matrix $X^{2}$ are $0$ and
$-\zeta^{2}$, and the {\it strong ellipticity condition} (2.45)
is {\it not fulfilled}, since, for strong ellipticity to
hold, the matrix $(X^{2}+ \II \zeta^{2})$ should be
strictly positive. This is why, for gravitational perturbations,
equation (2.32) {\it does not have a unique solution} for
$\lambda=0$, i.e. $\mu={\zeta}$, and any $\zeta$.
Technically, the lack of strong ellipticity implies that the
heat-kernel diagonal, although well defined, has a 
non-standard non-integrable behaviour as $r \rightarrow 0$.

\bigskip
\bigskip
\leftline{\sectionfnt 4. Concluding remarks}
\vskip 0.3cm
\noindent
Euclidean quantum gravity is an approach to the quantization
of the gravitational field that was stimulated by the need
to obtain a well defined path-integral representation of
out-in amplitudes. Although the main task remains too difficult,
since the gravitational action is unbounded from below [1, 2],
the Euclidean framework (more precisely, Riemannian) 
has led to rigorous results on the theory
of gravitational instantons (see [14] and papers therein), to
fascinating ideas in quantum cosmology [2, 14] and, more recently,
to a series of exciting developments on the subject of mixed 
boundary conditions in quantum field 
theory [2, 4--13, 15--18]. In
particular, it is by now clear that a fertile interplay exists
between the problems of spectral geometry [2, 12, 18] and the
Euclidean approach to quantum gravity and quantum cosmology.

Let us now discuss the meaning of our theorem 3.1 for 
Euclidean quantum gravity. As we have seen, for $\lambda=0$
the boundary conditions do not fix the solution in a unique
way. In other words, ``something wrong" occurs in the
zero-mode sector of the spectrum. Usually, for an elliptic
problem there are only a finite number of negative and zero-modes.
This leads in turn to a well known theorem about the standard
asymptotic behaviour of the heat kernel as $t \rightarrow
0^{+}$ [12]. When strong ellipticity is broken, however, there
can be {\it infinitely many} zero-modes; more generally, in the
neighbourhood of zero, the spectrum can be infinitely degenerate.
This is a highly undesirable property which leads to the
non-existence of the trace of the heat kernel, since the latter
includes summation over all modes. For the time being, the 
physical consequences remain unclear, at least to the authors.

Anyway, to obtain a meaningful formulation of Euclidean 
quantum gravity on manifolds with boundary, one has to
regularize the problem in such a way that the infinitely many
zero-modes do not appear. For example, to obtain a unique
solution one can introduce a regularization parameter, say
$w$, by rescaling
$$
\Gamma^{i} \rightarrow w \Gamma^{i} \; \; \; \;
Y^{i} \rightarrow w Y^{i} \; \; \; \; 
X \rightarrow w X
$$
where $w$ is a positive constant smaller than 1, and then
take the limit $w \rightarrow 1$ at the end of all
calculations. However, such a regularization would break
the gauge invariance, which was the initial motivation for the
consideration of generalized boundary conditions [2, 4--6].

It therefore seems that the analysis of Euclidean quantum 
gravity on manifolds with boundary faces a deep crisis: if
one avoids tangential derivatives in the boundary operator,
the resulting boundary conditions are not completely invariant
under infinitesimal diffeomorphisms [2, 5, 15]. On the other
hand, tangential derivatives in the boundary operator lead
to a boundary value problem which is not strongly elliptic,
as we have proved and emphasized. What should be checked is
whether the ghost fields, subject to the boundary conditions
(1.7), compensate exactly the effect resulting from infinitely
many zero-modes for gravitational perturbations. Ultimately,
however, a formulation should be achieved where both modes
(gravitational and ghost) are ruled by a strongly elliptic 
boundary value problem. Unless a way out to the 
dilemma is found which does not involve {\it ad hoc} 
assumptions, one should perhaps admit that the consideration
of boundaries is as essential as problematic in the attempts
to quantize the gravitational field in the Euclidean regime.

\bigskip
\bigskip
\leftline{\sectionfnt Acknowledgements}
\vskip 0.3cm
\noindent
We are grateful to Andrei Barvinsky, Stuart Dowker,
Alexander Kamenshchik, Klaus Kirsten and Hugh Osborn for
correspondence and conversations. The work of IA was
supported by the Deutsche Forschungsgemeinschaft.

\bigskip
\bigskip
\leftline{\sectionfnt Appendix}
\vglue0pt
\vskip 0.3cm
\vglue0pt
\noindent
\vglue0pt
This appendix describes some geometric constructions frequently
used in our paper. We consider a compact Riemannian manifold
$M$ of dimension $m$ with a positive-definite Riemannian metric $g$ 
and with a smooth boundary $\partial M$.
In the neighbourhood of $\partial M$
there exists a narrow strip, say $\Omega$, which is {\it locally}
a direct product
$$
\Omega=[0,\varepsilon] \times \partial M .
\eqno (A1)
$$
Following [12] we define a ``moved" boundary
$$
\partial M(r)= \left \{ x \in M: r(x)=r \right \}
\; \; \; \; r \in [0,\varepsilon] 
\eqno (A2)
$$
where $r(x)$ is the normal distance of a point $x$ to
$\partial M$. Thus, $\partial M(r)$ is a surface that is
parametrized by $r$ and coincides with $\partial M$ at
$r=0: \partial M(0)=\partial M$. This makes it possible to
obtain a natural foliation of $M$ in the neighbourhood of
its boundary.

We denote the local coordinates on $\partial M(r)$ and $\Omega$ by
${\hat x}^{i}$ ($i=1,...,m-1$) and $x^\mu=x^{\mu}(r,{\hat x}^{i})$
($\mu=1,\dots,m$), respectively.
The basis of vector fields in $T(\Omega)$ is 
$e_{a} \equiv (N,e_{i})$, with $a$ ranging from 1 through
$m$, where
$$
N\equiv\left|{\partial /\partial r}\right|^{-1}
{\partial\over \partial r}
\eqno (A3)
$$
is the unit normal vector field to $\partial M(r)$,
$|\partial/\partial r|^2=g(\partial_r,\partial_r)$, and $e_{i}\equiv
\partial / \partial {\hat x}^{i}$ is the basis of vector fields
in $T(\partial M(r))$.
The dual basis of 1-forms, say
$e^{a} \equiv (\omega,e^{i})$, is defined by
$$
<e^{i},e_{j}>=\delta_{j}^{i}
\qquad
<\omega,N>=1
\eqno (A4)
$$
$$
<\omega,e_i>=<e^i,N>=0.
\eqno (A5)
$$
The metric on  $\partial M(r)$ is defined by
$$
e_{i}\cdot e_{j}\equiv g(e_i,e_j)=\gamma_{ij}.
\eqno (A6)
$$
The normal covariant derivative is then $\nabla_{0} \equiv 
\nabla_{N}$, whilst tangential covariant derivatives are
$\nabla_{i} \equiv \nabla_{e_i}$.
The second fundamental form of $\partial M$ is defined by 
$$
\nabla_0 e_i=\nabla_i N=K_{i}^{j} e_{j}.
\eqno (A7)
$$
Last, the Levi-Civita connection $\hat\nabla$ on $\partial M(r)$ is
defined to be compatible with the metric $\gamma_{ij}$, i.e.
$
\hat\nabla_k\gamma_{ij}=0.
$

\bigskip
\bigskip
\leftline{\sectionfnt References}
\vskip 0.3cm
\item{[1]}
Hawking S W 1979 in {\it General Relativity, an Einstein
Centenary Survey}, eds. S W Hawking and W Israel 
(Cambridge: Cambridge University Press)
\item{[2]}
Esposito G, Kamenshchik A Yu and Pollifrone G 1997 
{\it Euclidean Quantum Gravity on Manifolds with Boundary}
({\it Fundamental Theories of Physics} {\bf 85})
(Dordrecht: Kluwer)
\item{[3]} 
DeWitt B S 1965 {\it Dynamical Theory of Groups and Fields}
(New York: Gordon and Breach) 
\item{[4]}
Barvinsky A O 1987 {\it Phys. Lett.} {\bf 195B} 344
\item{[5]}
Moss I G and Silva P J 1997 {\it Phys. Rev.} D 
{\bf 55} 1072
\item{[6]}
Avramidi I G, Esposito G and Kamenshchik A Yu 1996 
{\it Class. Quantum Grav.} {\bf 13} 2361
\item{[7]}
McAvity D M and Osborn H 1991 {\it Class. Quantum Grav.}
{\bf 8} 1445
\item{[8]}
Esposito G, Kamenshchik A Yu, Mishakov I V and Pollifrone G
1995 {\it Phys. Rev.} D {\bf 52} 3457
\item{[9]}
Avramidi I G and G. Esposito G 1997 ``New Invariants in the
One-Loop Divergences on Manifolds with Boundary"
(old version in hep-th/9701018).
\item{[10]}
Dowker J S and Kirsten K 1997 ``Heat-Kernel Coefficients 
for Oblique Boundary Conditions" (hep-th/9706129).
\item{[11]}
Abouelsaood A, Callan C G, Nappi C R and Yost S A 1987
{\it Nucl. Phys.} B {\bf 280} 599
\item{[12]}
Gilkey P B 1995 {\it Invariance Theory, the Heat Equation 
and the Atiyah-Singer Index Theorem} (Boca Raton, FL: 
Chemical Rubber Company)
\item{[13]} 
Desjardins S 1995 {\it Heat Content Asymptotics}
(Ph D Thesis, University of Oregon) 
\item{[14]}
Gibbons G W and Hawking S W 1993 {\it Euclidean Quantum 
Gravity} (Singapore: World Scientific)
\item{[15]}
Luckock H C 1991 {\it J. Math. Phys.} {\bf 32} 1755
\item{[16]}
Vassilevich D V 1995 {\it J. Math. Phys.} {\bf 36} 3174
\item{[17]}
Marachevsky V N and Vassilevich D V 1996 
{\it Class. Quantum Grav.} {\bf 13} 645
\item{[18]}
Esposito G 1997 ``Dirac Operator and Spectral Geometry"
(hep-th/9704016)

\bye